# First-principles computational methods for quantum defects in two-dimensional materials: A perspective

Hosung Seo[1,2,3,4]*, Viktor Ivády[5,6,7]*, and Yuan Ping[3,8,9]*

[1]SKKU Advanced Institute of Nanotechnology, Sungkyunkwan University, Suwon, Gyeonggi 16419, Republic of Korea

[2]Department of Physics and Department of Energy Systems Research, Ajou University, Suwon, Gyeonggi 16499, Republic of Korea

[3]Department of Materials Science and Engineering, University of Wisconsin-Madison, Madison, WI 53706, United States

[4]Center for Quantum Information, Korea Institute of Science and Technology, Seoul 02792, Republic of Korea

[5]Department of Physics of Complex Systems, Eötvös Loránd University, Egyetem tér 1-3, H-1053 Budapest, Hungary

[6]MTA–ELTE Lendület "Momentum" NewQubit Research Group, Pázmány Péter, Sétány 1/A, 1117 Budapest, Hungary.

[7]Department of Physics, Chemistry and Biology, Linköping University, SE-581 83 Linköping, Sweden

[8]Department of Chemistry, University of Wisconsin-Madison, Madison, WI 53706, United States

[9]Department of Physics, University of Wisconsin-Madison, Madison, WI 53706, United States

(*Authors to whom correspondence should be addressed: seo.hosung@skku.edu, ivady.viktor@ttk.elte.hu, yping3@wisc.edu)

**ABSTRACT**

Quantum defects are atomic defects in materials that provide resources to construct quantum information devices such as single-photon emitters (SPEs) and spin qubits. Recently, two-dimensional (2D) materials gained prominence as a host of quantum defects with many attractive features derived from their atomically thin and layered material formfactor. In this perspective, we discuss first-principles computational methods and challenges to predict the spin and electronic properties of quantum defects in 2D materials. We focus on the open quantum system nature of the defects and their interaction with external parameters such as electric field, magnetic field, and lattice strain. We also discuss how such prediction and understanding can be used to guide experimental studies, ranging from defect identification to tuning of their spin and optical properties. This perspective provides significant insights into the interplay between the defect, the host material, and the environment, which will be essential in the pursuit of ideal two-dimensional quantum defect platforms.

**OVERVIEW**

Deep-level defects in semiconductors are a leading platform for the development of innovative quantum information technologies in the solid state[1]. Several special defect complexes in diamond and SiC are front-running candidates for implementing large-scale quantum computation, non-destructive nano-scale quantum sensors, and quantum communication networks[2]. Attractive features of the defect systems as quantum hardware are mainly derived from unpaired electrons tightly localized at the defect site possessing several orbital states, which are well separated from the bulk states in energy[3,4]. Optical cycling can be generated between the orbital states emitting single photons at definite wavelengths, which constitute building blocks for quantum photonic circuits[5,6]. In special cases, the unpaired electrons have a non-zero ground-state spin that can be employed as a qubit[7]. These defect spin qubits can be optically initialized and read-out, and controlled by using microwaves, while maintaining their coherent quantum properties over millisecond to second time-scale[8,9]. Furthermore, the emitted single photons can be entangled with the spin, which can be used to build a spin-photon interface[10,11].

Very recently, the scope of quantum defects has further expanded since the discovery of single-photon emitters (SPEs) from two-dimensional (2D) materials such as hexagonal boron-nitride (h-BN) and transition metal dichalcogenide (TMD)[12–17]. Notably, several spin qubit candidates were recently realized in h-BN, including negatively charged boron vacancies ($V_B^-$)[18] and carbon-related defect centers[19–23]. Search of similar spin defects are also underway for TMD materials. Spin defect characteristics of carbon impurity and sulfur vacancies were identified in $WS_2$ and $MoS_2$, respectively[24,25]. Additionally, first-principles calculations predicted several promising defect candidates in h-BN and TMD as optically active spin qubits[26–31].

The 2D-materials-hosted quantum defects readily attracted a great amount of attention owing to their excellent formfactor[32]. The 2D nature of the host crystal may enable superior photon extraction efficiency, strong light-matter interaction, enhanced quantum coherence, and scalable fabrication that could outperform quantum defect systems embedded in conventional three-dimensional (3D) hosts such as diamond and SiC. In addition, these 2D materials provide rich possibilities for hetero-integration regardless of lattice matching conditions due to the weak van der Waals (vdW) binding energy between layers. Several rudimentary demonstrations were recently achieved in realizing such attractive features such as SPEs integrated with photonic and plasmonic devices[33,34], deterministic fabrications by using nanoindentation[35], and vdW heterostructure-based control of qubit properties[36–40]. The surface proximity of defect qubits in 2D materials can also enhance their sensitivity to external fields, making them excellent candidates for atomically thin quantum sensors[41–44]. We also remark that review articles are available for the recent experimental development made for 2D-hosted quantum defects[6,45–49].

It is essential to perceive that most of the challenges and advantages of 2D-hosted quantum defects[6,45–49] are derived from the defects being active open quantum systems, which interact with several environmental parameters. The coupling of spin defects to electric fields can be harnessed to tune their optical emission frequency and linewidth[50,51]. The coupling also allows for sensing electric and magnetic fields at nanoscale[52]. On the other hand, the same coupling mechanism also causes fast spin decoherence due to the coupling to magnetic and electric field noises in materials, which may limit the sensitivity of quantum sensors[53]. Interaction with the host's phonons may also significantly broaden the optical linewidth, leading to a limited number of coherent photons that can be extracted[54]. The coupling between SPEs and the charge fluctuation in the environment can induce significant spectral diffusion[55]. More importantly, these effects can be greater in 2D formfactor than in 3D formfactor[56], thus requiring an accurate understanding of these interactions and their effects on the defects.

In this perspective, we describe first-principles theories, which are capable of addressing the aforementioned issues for 2D-hosted quantum defects. Since the discovery of the 2D-hosted quantum defects, relevant theories have been rapidly developed. We found, however, that review and perspective articles on this subject are scarce in the literature. In our perspective, we focus on first-principles theories that can be performed without adjustable parameters, so that they can be applied to accurately predict the open quantum system phenomena starting from the microscopic constituents of a 2D-hosted quantum defect. In what follows, we discuss 1) spin decoherence and relaxation, 2) electronic and optical properties, and 3) defect engineering, mostly focusing on h-BN and TMD materials. In each section, we summarize the basic physics and necessary first-principles computational methods. Then, we present several recent examples of applications of the methods for quantum defects in 2D materials. Each section also provides an outlook for open questions, controversial issues, and potential future development necessary to address these issues.

## I. SPIN DECOHERENCE AND RELAXATION

### A. Spin environments of h-BN and TMD materials

A coherent superposition of a qubit's spin sublevels is an essential resource to implement almost any quantum information protocol. However, decoherence of spin qubits is inevitable due to the interaction with the surrounding environment. Due to the magnetic moment of the spin, one of the dominant decoherence sources is the magnetic field fluctuation present in the nuclear spin bath[57]. Compared to the nuclear spin bath of diamond and SiC[58,59], those of the 2D materials have several distinctive features. In h-BN, all the naturally occurring boron and nitrogen isotopes are spinful, leading to a very dense nuclear spin bath. Furthermore, some of the nuclear spins are larger than 1, leading to a large Hilbert space and more complex fine and hyperfine structures than those of diamond and SiC. In TMD, the nuclear spin bath depends on the elemental constituents of the material. In W-based TMD, the situation is similar to the diamond and SiC, where the nuclear spins are dilute and mostly ½. In Mo-based TMD, however, the nuclear spins could be as large as 5/2.

The 2D formfactor of h-BN and TMD also pass on several unique features in the spin-spin interactions, which play essential roles in both the qubit and the bath spin dynamics. For the hyperfine interaction, the pseudo-secular $A_{zx}$ and $A_{zy}$ terms can completely vanish in the qubit-containing plane if the spin density of the qubit keeps the mirror symmetry perpendicular to the *c* axis. In this case, the nuclear spins in the defect plane are protected from hyperfine-induced nuclear spin flips, which would be highly beneficial for nuclear spin memory applications[60]. The intrinsic nuclear dipole-dipole interaction is also largely constrained by the layered structure of the 2D hosts. Considering that the interlayer distance in 2D materials is typically larger than 3 angstroms due to the weak vdW interaction, strongly coupled nuclear spin clusters could likely form in a plane, while nuclear spins at different planes only interact with each other weakly. These characteristics of nuclear spin interaction were identified as a source of several distinct features of the quantum spin dynamics in 2D materials[61,62]. Ye et al. showed that isotopic purification can be much more effective in 2D than in 3D materials[61]. Onizhuk et al demonstrated that the decoherence of a spin qubit embedded in a low-dimensional TMDC heterostructure can undergo a transition from a classical to a quantum regime when the thickness of the host becomes thinner than 10 nm[62].

### B. Spin-bath driven decoherence: cluster correlation expansion

Qubit decoherence is in general a result of entanglement with the bath state due to qubit-bath interactions. Thus, accurate prediction of the decoherence requires computation of quantum evolution of the entire bath comprising of interacting many nuclear spins. Since the exact diagonalization of such a large system is intractable, various cluster expansion schemes were developed[63–66], from which the cluster correlation expansion (CCE) method[67,68] has recently stood out. In CCE, the contribution from the entire nuclear spin bath is divided into irreducible cluster correlation terms. Importantly, spin-pair contributions are shown to give numerically converged results for free induction decay (FID) and Hahn-echo decay in conventional 3D materials, making the qubit decoherence prediction more tractable. In the conventional CCE approach, secular approximation is normally adopted assuming that the qubit population is not altered during the decoherence process. This constraint can be relaxed in CCE by including the entire qubit Hamiltonian in the cluster expansion, leading to the generalized CCE (gCCE) method[69,70]. Using gCCE, one can compute more exotic decoherence phenomena in which non-secular hyperfine interaction plays an important role, such as clock transitions[71–73], decoherence at qubit level crossing points[69], and spin dynamics in qubit registers[74,75].

As spin decoherence is determined by the quantum bath dynamics, accurate prediction of spin Hamiltonian is necessary. For defect spins, these spin Hamiltonian tensors were largely governed by their ground-state spin density, thus the electronic structure, and the lattice relaxation, for which density functional theory (DFT) is a great tool[76]. Several important developments were made in the past decade, which include accurate calculations of the zero-field splitting tensor by including the spin-spin[77–79], spin-orbit[80], and spin contamination correction[81], real-space calculation schemes[82], and finite-size-effect free hyperfine calculation in DFT[83].

### C. Spin qubit dynamics in h-BN

The predictive power of the cluster expansion theories and DFT made significant contributions to the early stage of developing spin defects in 2D materials and understanding their spin coherent properties. The hyperfine interaction and the zero-field splitting of the ground-state spin of $V_B^-$ in h-BN were accurately computed by DFT, contributing to the defect identification[84,85] (Fig. 1a). The overall spin decoherence of spin qubits in h-BN was predicted even before the realization of $V_B^-$ spin qubits[61], and it was refined later by considering the DFT-computed spin density of the $V_B^-$ in h-BN[86,87]. Combining the experiment and gCCE theory, the spin decoherence time was shown to be limited to tens of ns even in a reasonably large magnetic field of around 100 G due to the dense h-BN nuclear spin bath[86] (Fig. 1b). Application of a magnetic field larger than 1 T was shown to be necessary to suppress the nuclear spin bath noise, which increases the $T_2$ time up to tens of microseconds[87], which was experimentally verified[88] (Fig. 1c). The contact hyperfine interaction and the nuclear quadrupole interaction were identified as key factors governing the decoherence in h-BN. The spin decoherence was predicted to be significantly suppressed in h-$^{10}$BN (Fig. 1b and 1d), which is mainly due to the lower gyromagnetic ratio of $^{10}$B than that of $^{11}$B despite its larger nuclear spin. Notably, the synthesis of isotopically engineered samples was recently made possible, and several recent experiments showed that the EPR linewidth of $V_B^-$ becomes much narrower in h-$^{10}$BN[89,90], confirming the previous theoretical prediction.

It is worth noting that nuclear spins-related effects may be utilized in different flavors of quantum technologies such as quantum memory[91], qubits[92], and quantum simulation[93]. In particular, the dense nuclear spin environment in h-BN can give rise to a plethora of exotic quantum phenomena such as non-Markovian dynamics[94]. It has been also demonstrated that Ramsey measurement of the $V_B^-$ center at moderate magnetic field values can provide unexpectedly long coherent oscillations exceeding the $T_2$ time measured in the same sample[95]. One possible explanation is the partial polarization and weak microwave drive of the nuclear spin bath through the mixture of the electron and nuclear spin states[95]. The strong hyperfine coupling can also be utilized to hyperpolarize the h-BN nuclear spins through the optical pumping of a $V_B^-$ ensemble[96,97]. Here nuclear spin is optically polarized and coherent control at room temperature through optically detected nuclear magnetic resonance (ODNMR). The hyperpolarization coupled with full quantum controls over the h-BN nuclear spins may lead to a 2D quantum spin simulation platform[98,99].

**D. Phonon-induced spin relaxation: mater-equation-based approach**

The other critical dynamical property of spin qubits is spin relaxation time ($T_1$), which quantifies the timescale of population decay of spin qubits, and also ultimately limits the spin decoherence time as $T_2 < 2T_1$. The $T_1$ process is primarily driven by spin-phonon coupling, which is mainly categorized into two: coupling through position-dependent spin-exchange interaction (between electron spins or electron spin-nuclear spin coupling) or coupling through spin-orbit interactions (often the case for single spin relaxation, or spin-1/2 defects). The former is more dominant in low spin-orbit systems, such as NV center in diamond, while the latter is more dominant in strong spin-orbit systems such as spin defects in ZnO. First-principles approaches based on the former mechanism describe reasonably well the $T_1$ time of NV center, through perturbation theory with Fermi's Golden rule[100] (Fig. 2a) or density-matrix perturbation theory[56], which was also applied to $V_B^-$ in h-BN. The spin-phonon interaction is computed through finite differences for derivatives of the interaction Hamiltonian along phonon eigenvectors. The latter mechanism is much less studied from first principles, most work has been performed from the effective mass approximation and deformation potential for electron-phonon coupling with parameterized model Hamiltonian. It has been shown to be effective in describing shallow impurities' spin relaxation in ZnO[101]. Note that for deep levels, the two-phonon Raman process or the Orbach process may be more dominant than the one-phonon process in shallow impurities. Such a higher-order phonon process for the spin-orbit mechanism of $T_1$ from *ab initio* has mostly been studied in molecular qubits[102]. There, a spin-vibrational orbit interaction was proposed for a better description of spin-orbit interaction in $T_1$. However, future development is still required to describe the spin-phonon relaxation time of deep defects in solids with strong spin-orbit couplings.

Such development may be built on recent work of *ab initio* spin relaxation and dephasing in solids[103–106], then generalized to spin defect states. In these studies, the authors developed density-matrix dynamics for coupled spin and carrier relaxation, with coherent and incoherent dynamics governed by the Lindblad master equation (Fig. 2b). Such formalism is derived from open quantum dynamics, starting from reduced one-particle electron density matrix and general electron and phonon Hamiltonians. By considering the phonon as the "bath" coupling with electronic systems, different approximations at Markovian and non-Markovian levels have been derived, depending on the necessary memory effect of the bath (or the feedback between systems and baths)[107]. In the context of spin dynamics, the coherent dynamics takes into account external magnetic field and light-matter interaction. In the incoherent dynamics, electron-phonon interaction with self-consistent spin-orbit coupling has described the spin relaxation of disparate systems well. Another type of scattering related to impurities and electron-electron interactions can be naturally included as well. For example, Eillot-Yafet and D'yakonov-Perel's mechanism has been naturally described, and the effect of inhomogeneous broadening such as $g$ factor variation on spin dephasing under external magnetic field has been accurately captured as well[104]. Experimentally, the spin relaxation mechanism can be determined by the distinct power-law of temperature and magnetic field dependence[101]. For higher-order phonon processes (e.g. Raman), the electron-phonon scattering matrix needs to be computed at the second-order perturbation theory.

Additionally, spin-orbit-induced spin relaxation could be quite dominant in transition metal dichalcogenides (TMDs) unlike h-BN, which could be investigated in the future. Currently, reliable experimental results on spin defects $T_1$ in TMD are still lacking for the benchmark of computational methods. Given the strong spin-valley locking effect for holes in TMD, the spin lifetime of holes has been extraordinarily long ($\mu$s), which was well reproduced from first-principles density-matrix dynamics[103]. How spins of defects in TMD interact with intrinsically "locked" spin due to strong spin-orbit could be an interesting question to answer. We also note that *ab initio* electron-phonon coupling for defect supercells is numerically challenging, considering the high energy resolution needed for accurate energy conservation description between spin states and phonon states. More efficient numerical methodology for electron-phonon coupling of defect supercells such as embedding methods or machine-learning methods may be necessary to be developed in the future.

E. Spin relaxation due to spin baths: extended Lindbladian method

When it comes to decoherence involving cross-relaxation with a bath spin, CCE methods may not be applicable. For such cases, an extended Lindbladian method has been suggested[108] building on the approximations already introduced in the gCCE method. In contrast to the CCE and gCCE methods, which calculate the time evolution of the off-diagonal element of the reduced density matrix, the extended Lindbladian method approximates the time evolution of the diagonal elements of the reduced density matrix of the central spin. To retain spin momentum conservation lows a coupled dynamics of few-spin-systems is computed through extended and coupled Lindbladian master equations[108]. This method has already been utilized to calculate magnetic field dependence of the spin relaxation time $T_1$ and cross-relaxation features between a defect qubit and surrounding spin bath in diamond[108] and SiC[109–111] (Fig. 2c).

The theory of spin relaxation and cross-relaxation effects is yet to be studied in detail in h-BN and TMDs. Spin relaxation due to the spin bath is expected to be a dominant source at low temperatures. The fine and hyperfine structure-dependent relaxation phenomena can provide important information on the composition of the spin bath and may be utilized, e.g. to determine the types and concentration of the spins in the local environment of the defects. Relying on robust lifetime measurements of spin qubits in thin 2D materials can provide information not only on the host's spin bath but also on the magnetic environment of the layers. Such relaxometry applications possess a high potential for sensing and characterization techniques, which motivates further studies in this direction.

## II. ELECTRONIC AND OPTICAL PROPERTIES

### A. Density functional theory (DFT) methods

The workhorse for computing the physical properties of defects in solids is DFT[112]. Building upon the long history of defect calculations using DFT[113], DFT-based theories for quantum defects were also significantly advanced in the past two decades and become indispensable tools for quantum defect research[114]. One of the most important quantities for a quantum defect is the defect formation energy (DFE) as a function of the Fermi level, from which one can extract critical information on thermodynamic stability, charge transition levels, ionization energy, and photostability[112,113]. To compute the DFE of charged defects in 2D materials, it is important to apply a proper charge correction scheme, which is available in the literature[115]. Another key ground-state property of a quantum defect is the local defect geometry as it not only determines the symmetry of the defect but is also the foundation for any other advanced calculations[116,117]. The defect geometry also determines the electric dipole, which affects the electric field sensitivity via the Stark effect[118]. A high-symmetry defect would be also desirable for realizing a spin-photon interface[10,119]. The single-particle orbitals obtained from the ground-state calculations are used to quickly determine possible optical transitions in a defect and serve basis functions for more accurate and advanced many-body calculations[120].

Many of the key optical processes of a quantum defect heavily rely on the excited-state structure and its dynamics such as the optical initialization of spin qubits, generation of coherent single photons, and creation of a spin-photon interface. The accurate prediction of the excited-state properties, however, is one of the most challenging calculations on which lots of previous efforts were focused. One could start excited-state calculations by constrained DFT calculations with occupation modified between defect orbitals obtained compared to the ground state[121,122]. This technique was shown to work well for defects in 3D materials[123,124] (Fig. 3a), making it also a widely adopted method for defects in 2D crystals[119,125–127]. Constrained DFT is computationally less costly and particularly convenient for total energy and force calculations than other excited-state methods. However, given it's a mean-field level theory, it suffers both theoretical and computational challenges for describing excited states, leading to continuous recent efforts to better describe excited states of quantum defects in materials.

### B. Excited states of quantum defects: beyond DFT

It should be noted that ultrathin 2D semiconductors/insulators inherently have weak and anisotropic dielectric screening, which leads to strong electron correlation and electron-hole interactions, thus requiring proper treatment of such effects in their electronic and optical properties[128–130]. The excitonic effect was the order of magnitude larger in 2D systems than in 3D systems, which introduced stronger interactions of defects with other quasiparticles, such as exciton-defect and defect-exciton-phonon interactions[131]. This also causes significant differences between electronic and optical excitation energies (e.g. differences between quasiparticle gaps and optical gaps), which cannot be captured correctly by mean-field theory including DFT.

Several groups advanced the excited-state calculations of defects in 2D materials based on many-body perturbation theories such as GW-Bethe-Salpeter Equation (BSE)[127,132–135] (Fig. 3b). Many-body perturbation theory calculations normally start from inputs of DFT single-particle energies and wavefunctions, then compute quasiparticle energies at GW approximation (with self-energy as the convolution of one particle Green's function (G) and dynamically screened Coulomb interaction (W))[136,137]. Such a level of theory describes well the electronic band gaps or quasiparticle energies of a wide range of semiconductors and insulators. For optical properties, one needs to solve the BSE for neutral excitation, where the BSE kernel contains an exchange term (originated from the Hartree potential) and a direct term built with statically screened exchange interactions. Such approximations have been shown effective for describing optical properties in particular excitonic effects in solids[138,139]. For certain cases, for example, where double or higher excitations are important, dynamic screening in BSE would be necessary[140,141]; for optical properties of most two-dimensional semiconductors and their defects, static screening is adequate for accurate excitation energies and spectroscopic signatures in single excitations[130,142]. Further development of spin-flip BSE provides access to multi-reference states, beyond single Slater determinants in traditional BSE[143].

To push the accuracy of the excited state calculations further, in particular the inclusion of multireference states, theories describing accurately both static and dynamic correlations were developed such as quantum defect embedding theory (QDET)[117,144–146], CI-RPA embedding method[116], and density matrix renormalization group theory[147]. QDET as a tool for solid-state defect calculations can describe defects, host materials, and the interaction between defect and host accurately through the idea of quantum embedding. In particular, the active space orbitals that describe the defect related states are treated at an exact diagonalization level (full Configuration Interaction), which in theory is exact; the environment is treated at the GW approximation[148], which also described properly the dielectric screening and electron correlation; then the double-counting interaction at defect states was carefully treated. Double counting terms in the effective Hamiltonian of the active regions are terms that are computed both at the level of theory chosen for the active region and at the lower level chosen for the environment. Hence corrections (often called double counting corrections) are required to restore the accuracy of the effective Hamiltonian. In the original formulation of QDET presented in Ref.[144], they adopted an approximate double-counting correction based on the Hartree-Fock theory. In Ref.[148], they present a more rigorous derivation of QDET based on Green's functions where the authors derive a double counting correction that is exact within the $G_0W_0$ approximation and when retardation effects are neglected.

One important note is that excited-state forces remain challenging among all these advanced methods. Spin-flip time-dependent DFT (TDDFT) recently gained significance in quantum defect calculations and several recent developments[149–151] were made to compute the excited-state forces and geometry relaxation, spin-flip excitations, and the reliable prediction of the luminescence lineshape property. This method has the advantage of capturing multi-reference states and is computationally efficient for excited-state force calculations. While TDDFT with adiabatic approximation does not describe excitons in 2D materials, the inclusion of a nonlocal exchange-correlation kernel could partially address the issue here[152]. Theoretically, spin-flip TDDFT does not capture double excitation as well, which is captured in QDET or multireference wavefunction methods. Excited forces from spin-flip BSE have been recently developed, which include excitonic effects[143].

Recently, wavefunctions-based quantum chemistry (QC) methods have also gained visibility in the computational physics community. In contrast to periodic methods, QC models of point defects often consist of defected 2D flakes terminated with hydrogen. Due to the weak dispersive interaction between the layers and the highly anisotropic screening of the electronic repulsion, the use of a single-layer flake model may provide a reasonable approximation to 2D systems[30,153,154]. The moderate computational cost enables the use of advanced quantum chemistry tools, such as NEVPT2, CCSD, CCSD(T), and MRCI to mention a few[30,153]. QC software mostly uses atomic basis sets, whose convergence needs to be religiously checked, especially for vacancy-related defects[147,154]. The availability of accurate *ab initio* results could also facilitate the benchmarking and evaluation of DFT calculations for 2D point defect systems and help advance the theory of quantum defects by comparing DFT-based methods and the QC methods for excited-state properties[155]. Recent developments in several groups provide quantum chemistry methods for periodic solid-state systems, particularly for defect supercells, such as EOM-CCSD[156] and Particle-Particle Random Phase Approximation (PPRPA)[157]. These methods could provide promising avenues for self-consistent theory with a strong correlation in solids.

### C. Understanding the electronic environments

Quantum defects could exhibit enhanced light-defect interaction in 2D hosts, leading to increased quantum efficiency and strong exciton formation, which is largely attributed to the reduced dielectric screening in 2D[128,134,142,158] (Fig. 4a). In addition, the optical process occurring in a defect can be highly sensitive to the sharp interface in 2D van der Waals heterostructure, providing various engineering opportunities by choosing the surrounding environment such as the interface, the surface, and the substrate. It was demonstrated that color centers in h-BN can be activated and controlled by twisting the layers[38]. Moiré trapping potentials were shown to be an effective way to generate a quantum light source in TMDC heterostructures[159]. The effect of a dielectric environment such as substrates for defects in 2D materials has been studied in recent work[160–162] (Fig. 4b). Through the method of summation of dielectric polarizability of systems and substrates with reciprocal space linear interpolation technique, many-body perturbation theory can be performed at 2D interfaces, with their natural lattice constants. We found that the optical excitation energy or ZPL has a minimum change in the presence of different substrate screening, which is a direct result of the cancelation of the dielectric screening effect on quasiparticle gaps and exciton binding energy. However, we emphasize that this cancelation may not be always the case for any quantum defects in 2D materials and one should carefully examine the substrate effect on the optical excitation, which was shown to be significant in 2D materials[163].

One important factor for photonic applications is the defect's internal quantum efficiency, which is governed by the competition of radiative and nonradiative decay processes from the optically excited state. The quantum efficiency can be measured experimentally and combining with observed photoluminescence lifetime gives the components of radiative and nonradiative recombination time. The radiative lifetime of the excited state of a defect can be obtained from the Fermi's Golden rule with inputs of transition dipole moment and excitation energy from appropriate electronic structure theory. For defects in 2D materials, strong exciton-defect interaction requires methods taking into account excitonic effect, such as solving the BSE[127,134,164]. The nonradiative recombination rate of defects[112,165] has been calculated several times with static coupling approximation and effective phonon approximation for 3D systems[166–169] and 2D systems[170,171] (Fig. 4c). Comparison with experimental nonradiative capture coefficients has been performed for defects in GaN[166]. The complexity comes in when the excited-state potential surface is complicated, for example, involving dynamical and pseudo-Jahn Teller distortions that mix multiple excited states, where single particle theory is unable to describe electron-phonon coupling in the excited states.

Calculation of the rate of nonradiative processes is typically a challenging part, which demands high precision energy spectra, faithful characterization of local vibrational modes, accurate calculation of spin-orbit matrix elements, and often the inclusion of electron-phonon interaction beyond the Born-Oppenheimer approximation. Among non-radiative processes, one can distinguish between spin-conserving and spin non-conserving processes. The former is often termed as internal conversion while the latter as inter-system crossing. Internal conversion is mostly dominated by phonon-assisted nonradiative recombination. Such processes have been computed previously in the context of nonradiative capture coefficients for defects in semiconductor for 3D systems[166] and 2D systems[170]. It is worth noting that internal conversion can be neglected for visible emitters in h-BN, however, for near-infrared emitters this effect needs to be taken into account. The intersystem crossing is typically relevant for spin-carrying defects, especially for high spin defects, such as the $V_B^-$ in h-BN. Given the low quantum yield of $V_B^-$ in h-BN, such nonradiative process is important to consider.

Surprisingly, the layered formfactor of the 2D hosts could also allow one to design quantum emitters that are mechanically decoupled from the solid-state environment even at room temperature[54]. Subsequent investigation on the electron-phonon interaction showed that the emitter is largely insensitive to in-plane acoustic phonon modes, suggesting that the defect's active orbitals extend out-of-plane[172]. It was also suggested that an out-of-plane distorted defect structure[36] could be a strong candidate to explain the observed phenomena. Recently, however, DFT calculations were performed for in-plane $C_2C_N$ and out-of-plane-distorted $V_NN_B$, finding no indication that out-of-plane transition dipole is sufficient to provide lifetime-limited photons at room temperature[173]. So, the issue remains open, motivating further *ab-initio* investigations for finding different types of 'out-of-plane' defect candidates[125,174,175] with a reduced phonon-induced ZPL dephasing property. In addition, we note that the *ab-initio* electron-phonon matrix element calculations based on density functional perturbation theory[176] are generally too heavy to perform for a broad range of defect candidates. An efficient way to screen the defect candidates may be employing an effective theory based on ZPL-strain susceptibility[177], which can be computed efficiently by using DFT. A combination of the discovery of a wide variety of out-of-plane quantum defects in h-BN and the *ab-initio*-based effective theory would help identify the mechanism of the unique property of quantum emitters in 2D materials.

In addition to the homogeneous broadening of the ZPL, quantum emitters in 2D materials also suffer inhomogeneous broadening of their emission lines, which is also referred to as spectral diffusion. Previous studies revealed that one of the major sources of spectral diffusion is the Stark coupling of defect emission to fluctuating charge environment[51] as in 3D materials[50]. Theoretical understanding of the spectral diffusion in 2D materials, however, remains elusive, which requires the identification of ground-state and excited-state electric dipoles and the charge environment of the materials. The former relates to defect identification, or at least possible classification and survey of defect candidates in terms of Stark coupling property. The latter requires an in-depth study of common charged defects in the bulk and the surface, and their charge dynamics under relevant experimental conditions such as laser illumination or electric-field application. To understand the spectral diffusion, it is also necessary to compute and characterize the spectral function of the charge noise. We note, however, that first-principles methods for such characterizations for quantum defects have not been developed yet, while relevant studies have been performed in the trapped ion literature[178].

## III. Defect engineering

### A. Defect identification

Since the discovery of quantum defects in 2D materials, one of the literature's key issues is identifying their chemical composition and structure[12]. The use of electronic structure calculation is inevitable in this process today[84,127,179]. Solids can host zillions of different structural defects. The number of possible structures can be narrowed down by intuition, e.g. by taking into consideration the circumstances of creation and symmetry, and by carrying out high throughput calculations on a lower level of theories[180–182].

Defects that exhibit both a characteristic magnetic resonance signal and an optical signal can often be identified straightforwardly. The ground state fine and hyperfine structure of the defects serves as a fingerprint, which can be reliably calculated using hybrid DFT methods[81,83,183]. In particular, the hyperfine splitting of the nuclear spin states can now be predicted with a remarkable, few percent relative error[83] making the identification undoubtable. This led to the rapid assignment of the $V_B^-$ center and the corresponding ODMR signal[18,84,85]. Magnetic-field-dependent ODMR contrast determines the fidelity of optical readout and spin polarization. The theoretical prediction of ODMR contrast is challenging, and requires a precise description of electron-phonon coupling, spin-orbit coupling, and excited-state kinetics for strongly correlated excited states. Recently first-principles theory of ODMR contrast has been explored with kinetic master equation and excited-state relaxation rates from Fermi's

Golden rule[184]. Such calculations shine a light on future theory predictions of better defect candidates with efficient and reliable optical readout.

The identification is more challenging when the defect has no characteristic ODMR signal. This is the case for the unidentified ODMR-active carbon-related defects in h-BN[19–23] and for most of the visible emitters in h-BN[12,185–187] and TMDs[188–192]. Excited-state calculations play critical roles in identifying these defects by predicting the luminescence lineshape and the optical properties of various defect models that can be directly compared to experimental measurements. Similarly to the ODMR contrast, the features of the optical signal, e.g. polarization, Huang-Rhys, and Debye-Waller factor, and the frequency of dominant phonons coupled to the optical transition are good identifiers of the microscopic structure of the emitters. For instance, polarization data reflects the symmetry[120] of the defect and can be reliably obtained in DFT-based delta-SCF calculations[193] when the excited state has a single slater determinant character. For quantum defects in TMDC materials, however, defect-bound excitons may play an important role in optical processes, thus requiring the excitonic effect to be included in calculations[142,194]. Characteristic phonon energies carry information on the bonding of the atoms displaced upon optical excitation. Using large-scale, highly convergent calculations, the photoluminescence phonon sideband spectrum of defects can also be accurately calculated with DFT in 2D and 3D semiconductors[124,127,179,195,196] (Fig. 3a and Fig. 5a.) Calculations of photoluminescence spectrum beyond phonon sidebands and Frank-Condon approximations requires exciton-phonon coupling with a quasi-equilibrium treatment or a non-equilibrium treatment[197].

Another complication of defect identification in 2D materials comes from the sensitivity of quantum defects to strain that can be highly inhomogeneous in 2D materials[198]. It was previously shown that emission and ODMR features change dramatically defect-to-defect, which was attributed to the defect-strain coupling[185,199,200]. Several theoretical studies supported it by showing that the optical energy of various SPE candidates in h-BN is highly sensitive to lattice strain in h-BN[185,199,201,202]. Thus, to advance defect identification in 2D materials, it is necessary to develop experimental ways to characterize atomic-scale strain profiles around an individual defect[203] and to theoretically analyze the full strain susceptibility of a broad range of quantum defects under arbitrary and anisotropic strain[203].

### B. Deterministic generation of defects

The defect identification process may be accelerated if one knows better how defect forms[204] in 2D materials platforms at the atomic scale. Recently, various techniques were developed to create quantum defects in 2D materials such as irradiation with electrons[186,187], ions[205], and femtosecond laser[206] followed by high-temperature annealing. These methods were proven to be effective in deterministically generating emitters with well-defined emission frequencies with high spatial resolution. In each process, however, the defect formation would occur in different ways, and it should be also significantly affected by the distinctive 2D materials formfactor[207,208] and the lattice strain[209]. A microscopic understanding of the defect generation can be achieved by using first-principles molecular dynamics (FPMD)[210]. One of the main challenges is, however, that the activation energies of defect migration in materials can be as large as several electronvolts, thus limiting the direct applicability of FPMD simulations for the defect generation process. Several recent progresses were made in the literature, including the coupling of FPMD simulations with neural-network-based enhanced sampling techniques[211]. Continuous development of such computational methods will certainly benefit the rational design of quantum defects in 2D materials combined with the relevant experimental efforts.

Another promising avenue for deterministic generation of quantum defects in 2D materials is to use organic molecules absorbed on 2D-materials surface[212–214]. From a computational perspective, this approach brings a wide range of research opportunities and challenges. Pioneering theoretical studies are available in the literature, which focused on the atomic arrangement and optical properties of organic adsorbate-surface systems[215–217]. More studies are anticipated in the future, from a quantum defects perspective.

### C. Defect emission control

Implementation of quantum information protocols involving single photon emitters often requires that all emitted photons are identical in all possible degrees of freedom, such as frequency, polarization, and spectral shape[218]. This is, however, generally impossible for quantum defects as-is since the degrees of freedom become disordered due to the interaction with the solid-state environment. Thus, control over their optical properties is necessary and it can be achieved by coupling the defect to external parameters. Electric-field-driven control and strain engineering are often highly desirable. Electric-field-driven controls are important tools as they enable local control, which is beneficial to realize chip-scale integration[37,219]. Strain, which is intertwined with electric field in crystals, is also widely used to control the resonance frequency and the dipole (Fig. 5b). Compared to 3D hosts, electric field control and strain engineering in 2D materials would not only be effective but could provide extra opportunities. Considering a defect embedded in a thin 2D layer, a very high electric field could be applied, allowing for a large Stark shift[39,219]. It was also shown that 2D materials are ideal platforms for strain engineering due to their incredible mechanical flexibility and robustness[220,221]. In addition, highly inhomogeneous strain can be applied by nanoindentation, folding, or substrate engineering, which provides versatile tools to control the spin and optical properties of quantum defects[207,208,222,223].

The electric tunability of the ZPL via the DC Stark effect was demonstrated in 2018 for atomic defects in h-BN by exploiting vertical[36,224] and lateral[225] heterostructures and for SPEs in TMDC materials[39,226,227]. In addition, defects with V-shaped and quadratic-shaped Stark shifts were also observed[36,119]. While a few defect models in h-BN (Fig. 5c and Fig. 5d)[36,119] and in $WSe_2$[194] were proposed, many aspects of the Stark shift, however, remain unclear. From the theoretical perspective, the Stark shift is not straightforward to simulate using periodic DFT calculations as it involves a very accurate estimation of the dipole moment difference between the defect's ground state and excited state in the presence of an electric field. The delta-SCF method is normally adopted to describe the excited state of defects in h-BN[36,119]. However, the method may not apply to SPEs in TMDC materials as they likely involve defect-bound excitonic states[194]. A straightforward method to simulate the presence of an electric field is the saw-tooth waveform method, but this is only applicable for vertical Stark shifts for 2D materials[36,228]. One could build a hydrogen-terminated flake model for 2D materials to simulate the lateral Stark shifts, but the boundary effect may be hard to estimate. Estimation of the dipole moment difference for a charged defect should be also performed with great care[229]. In sum, the Stark shift behavior of quantum defects in 2D materials is still largely unexplored and their many aspects remain unclear, motivating extensive first-principles studies for a broad range of SPE candidates in 2D materials.

Strain effects on quantum defects in 2D materials can be directly estimated by using first-principles electronic structure theories[126,160]. Optical properties of quantum defects in h-BN and TMDs are largely tunable by strain with highly anisotropic response[185,189,222,224]. Depending on the symmetry of the defects and the nature of frontier orbitals, compressive or tensile strain can cause systematic blue or red shift of zero phonon line or change of radiative lifetime, qualitatively explained by molecular orbital theory[160] (Fig. 5e). Note that strain can also tune strongly the nonradiative lifetime of defects in h-BN, which we show strong strain dependence perpendicular to the $C_{2v}$ symmetry axis of $N_BV_N$ defect, but weak dependence parallel to symmetry axis. Such anisotropic strain dependence is largely determined by chemical bonding direction if strain directly affects the defect atomic bond length[170].

## CONCLUSION

To conclude, we showed that the use of first-principles simulations plays a crucial role in advancing quantum defects in 2D materials for both fundamental science and their practical applications. For spin decoherence and relaxation, we emphasized that the treatment of the qubit and the bath environment on the same footing is crucial for accurate prediction and understanding. We discussed how both DFT and beyond-DFT simulations can be utilized to investigate the key electronic and optical properties of quantum defects. We also discussed how such microscopic understanding is necessary to develop ways to manipulate their optical and electronic properties. Finally, we remarked that many issues remained to be addressed, such as spectral diffusion and defect identification. We believe, however, that continuous development and refinement of first-principles computational methods will play an essential role in creating active collaboration between theory and experiment, addressing the issues, and pushing the boundary of quantum defects in 2D materials toward the realization of robust solid-state quantum technologies.

## ACKNOWLEDGEMENT

This study was supported by the National Research Foundation (NRF) of Korea grant funded by the Korean government (MSIT) (No. 2023R1A2C1006270, RS-2024-00399417), by Creation of the Quantum Information Science R&D Ecosystem (Grant No. 2022M3H3A106307411) through the NRF of Korea funded by the Korea government (MSIT), and by Global - Learning & Academic research institution for Master's·PhD students, and Postdocs (G-LAMP) Program of the NRF of Korea grant funded by the Ministry of Education (No. RS-2023-00285390). This material is based upon work supported in part by the KIST institutional program (Project. No. 2E32971). This work was supported by Institute of Information & communications Technology Planning & Evaluation (IITP) grant funded by the Korea government (MSIT) (No.2022-0-01026). V.I. was supported by the National Research, Development and Innovation Office of Hungary (NKFIH) within the Quantum Information National Laboratory of Hungary (Grant No. 2022-2.1.1-NL-2022-00004) and within projects FK 135496 and FK 145395. V.I. also acknowledges support from the Knut and Alice Wallenberg Foundation through the WBSQD2 project (Grant No. 2018.0071). Y.P. was supported by the National Science Foundation under grant no. DMR-2143233.

# FIGURES

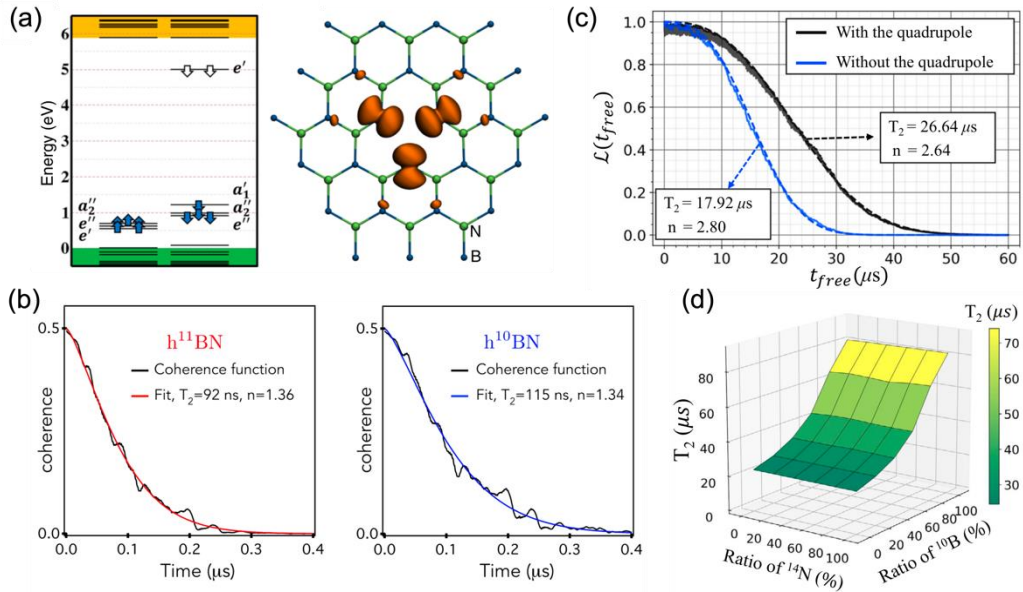

**Figure 1. Spin decoherence of the $V_B^-$ spin in h-BN. (a)** Single-particle electronic structure (left) and spin density of $V_B^-$ in h-BN as obtained by spin-polarized HSE06 DFT. Adapted from Ref.[84] under CC BY 4.0 license. **(b)** Simulated spin echo decay curve and corresponding stretched exponential fit for the $V_B^-$ defect in h$^{11}$BN (left) and h$^{10}$BN (right) in an applied magnetic field of 150 G. Adapted from Ref.[86] under CC BY 4.0 license. **(c)** Spin coherence of the $V_B^-$ spin in h-BN (black line) in an external magnetic field of 3 T. A hypothetical model of the bath in which the nuclear quadrupole interaction is ignored was considered for comparison (blue line). **(d)** Computed $T_2$ of the $V_B^-$ spin in h-BN as a function of the ratio of $^{14}$N and $^{10}$B nuclei in the lattice. An external magnetic field of 3 T is applied. It is worth noting that the suppression of the decoherence effect due to the isotopic purification in h-BN strongly depends on the strength of the external magnetic field. (c) and (d) are adapted from Ref.[87] under CC BY 4.0 license.

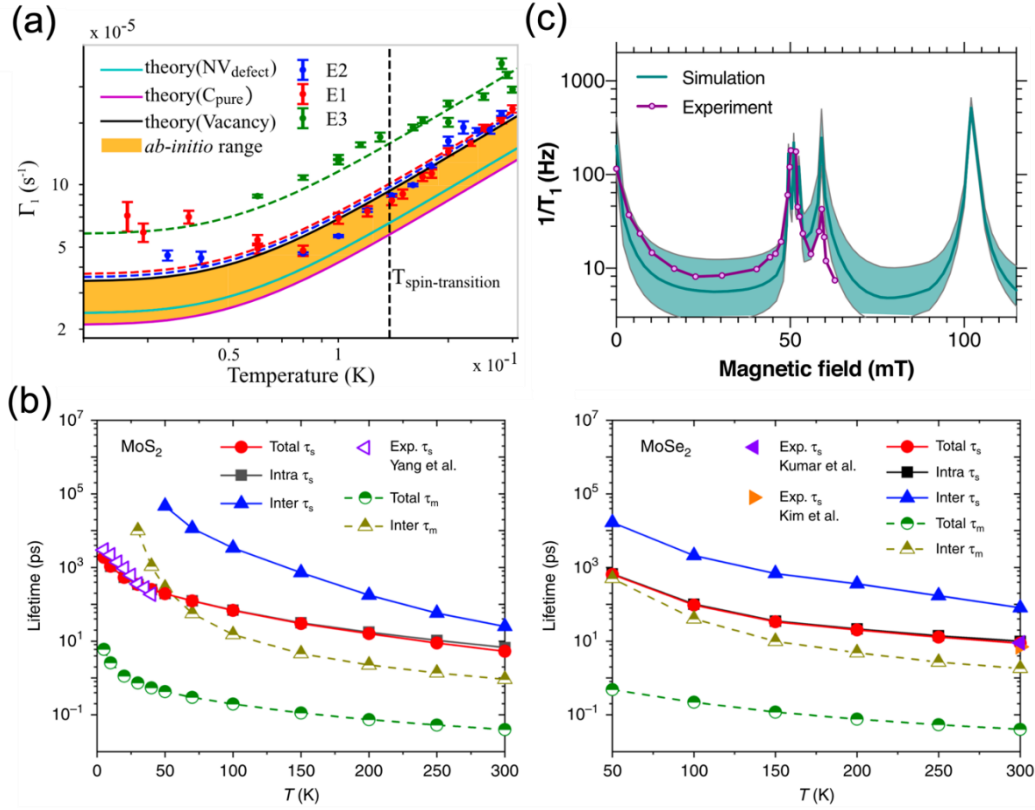

**Figure 2. Spin relaxation in 3D and 2D materials. (a)** The measured spin lattice relaxation rates for three different diamond NV samples (E1, E2, and E3; data taken from Ref.[230]). Dashed lines are least-square fits to the data for the temperature dependence. The theoretical results (yellow range) depend on the phononic density of states. Adapted from Ref.[100] under CC BY 4.0 license. **(b)** Predicted spin ($\tau_s$) and momentum ($\tau_m$) relaxation times of conduction electrons in $MoS_2$ (left) and $MoSe_2$ (right) with carrier concentrations of $5.2\times10^{12}$ cm$^{-2}$ and $5.0\times10^{11}$ cm$^{-2}$, respectively. Experimental data are taken from Ref.[231–234]. "Intra" and "inter" denote intravalley (within K or K′) and intervalley (between K and K′) scattering contributions to the relaxation times; intravalley processes dominate spin relaxation at and below room temperature. Adapted from Ref.[103] under CC BY 4.0 license. **(c)** Comparison between theoretical spin relaxation rate ($1/T_1$) and the experimental rate reported for diamond NV sample S2 at 20 K in Ref.[235]. Adapted from Ref.[108] under CC BY 4.0 license.

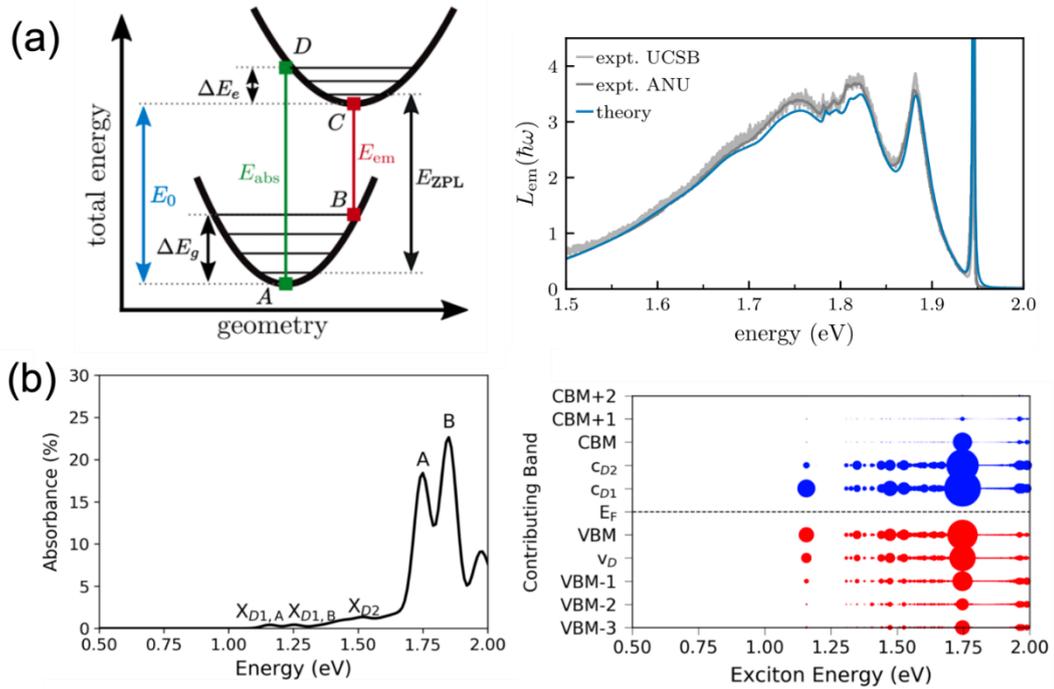

**Figure 3. Excited states of quantum defects. (a)** ) (left) One-dimensional representation of adiabatic potential energy surfaces in the ground and the excited states of diamond NV. $\Delta E_g$ and $\Delta E_e$ are lattice relaxation energies. $E_{ZPL}$ is the energy of the zero-phonon line. (right) Calculated normalized luminescence lineshape, compared with the experimental lineshape. Reproduced with permission from Razinkovas *et al.* Phys. Rev. B **104**, 045303 (2021)[124]. Copyright 2021, American Physical Society. **(b)** (left) Computed absorbance spectrum of 5×5 supercells of $MoSe_2$ with a single chalcogen vacancy. (right) Contributions of each band to each exciton state, plotted with respect to the exciton excitation energy (Only the A series of spin-orbit split excitons is included for clarity). The bands are labeled as either bulklike dispersive bands relative to the valence band maximum (VBM) and conduction band minimum (CBM) or as flat defect bands $v_D$, $c_{D1}$, and $c_{D2}$, with the occupied band contributions in red and unoccupied band contributions in blue. The size of each dot is proportional to square of the k-space electron-hole amplitude of the contribution from each band weighted by the oscillator strength of the exciton state. Reproduced with permission from Refaely-Abramson *et al.*, Phys. Rev. Lett. **121**, 167402 (2018)[135]. Copyright 2018, American Physical Society.

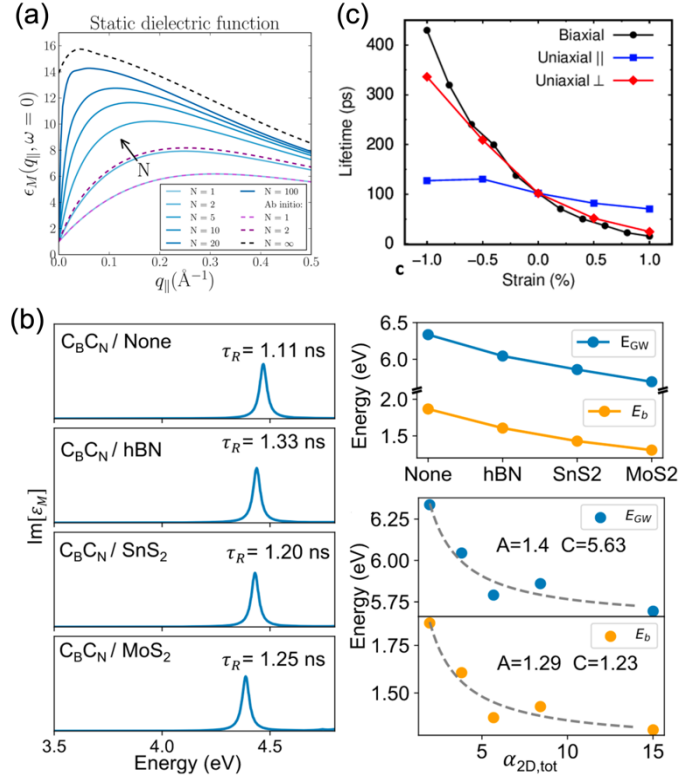

**Figure 4. Electronic environments of quantum defects in 2D materials. (a)** The macroscopic static dielectric function $\epsilon_M(q_\parallel, \omega = 0)$ of MoS$_2$ computed as a function of the in-plane momentum transfer for different number of layers, $N$. The dielectric functions increase monotonically with $N$ converging slowly toward the dielectric function of bulk MoS$_2$. Reproduced with permission from Andersen *et al.*, Nano Lett. **15**, 4616–4621 (2015)[128]. Copyright 2015, American Chemical Society. **(b)** (left) The BSE optical spectra of the C$_B$C$_N$ defect in monolayer h-BN on various substrates. (right top) The exciton binding energy and GW energy gap of the C$_B$C$_N$ defect in monolayer h-BN on different substrates. (right bottom) The GW energy and exciton binding energy as a function of total 2D polarizability. The grey dashed line is the fitting result. Reproduced with permission from Zhang *et al.*, 2D Mater. **10**, 035036 (2023)[160]. Copyright 2023, IOP publishing. **(c)** Strain-induced properties related to non-radiative recombination lifetime of the $^1B_1 - {^2B_1}$ defect-defect transition of N$_B$V$_N$ in monolayer h-BN. Reproduced with permission from Wu *et al.*, Phys. Rev. B **100**, 081407 (2019)[170]. Copyright 2019, American Physical Society.

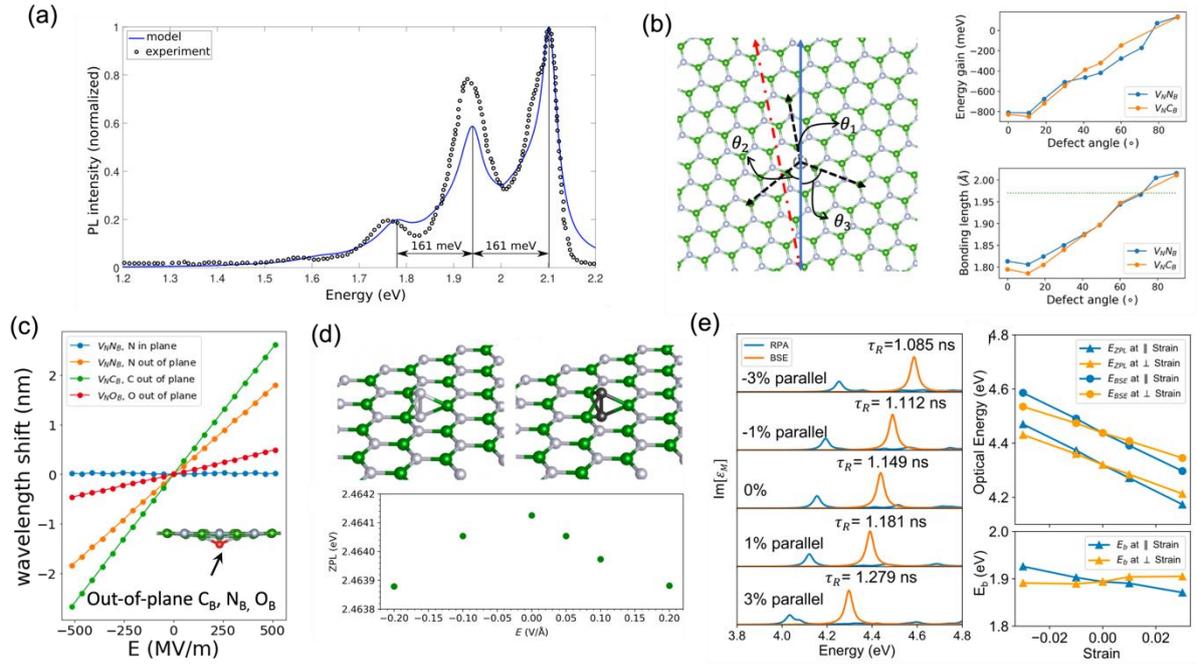

**Figure 5. Defect identification and control in h-BN. (a)** The calculated photoluminescence spectrum of $C_2C_N$ defect (inset) in h-BN, together with a typical experimental spectrum taken from Ref.[236]. The computed ZPL is blue-shifted by ∼0.5 eV to match the experimental ZPL value. Reproduced with permission from Jara *et al.*, J. Phys. Chem. A **125**, 1325–1335 (2021)[179]. Copyright 2021, American Chemical Society. **(b)** Alignment of the $V_NX_B$-type defects in h-BN wrinkles with the wrinkle direction. (left) Three possible defect angles ($\theta_1, \theta_2, \theta_3$) between the defect axis (black dashed arrows) and the wrinkle direction indicated by a solid arrow. (right top) The energy gain of the $V_NN_B$ and $V_NC_B$ defects as a function of the defect angle. (right bottom) The dimer length of the $V_NN_B$ and $V_NC_B$ defects as a function of the defect angle. Reproduced with permission from Yim *et al.*, ACS Appl. Mater. Interfaces **12**, 36362 (2020)[207]. Copyright 2020, American Chemical Society. **(c)** Theoretical ZPL shifts of neutral $V_NX_B$ defects (X = C, N, or O) in h-BN as a function of out-of-plane electric field. Inset shows the impurity atom ($X_B$) displaced out-of-plane, leading to the defect's symmetry being reduced to $C_{1h}$. Reproduced with permission from Noh *et al.*, Nano Lett. **18**, 4710 (2018)[36]. Copyright 2018, American Chemical Society. **(d)** Representation of (top left) nitrogen split interstitial defect and (top right) $C_N^2$ defect in the h-BN lattice. (bottom) The electric field dependence of the ZPL energy of the nitrogen split interstitial defect as obtained from first-principles calculations. Reproduced with permission from Zhigulin *et al.*, Phys. Rev. Applied **19**, 044011 (2023)[119]. Copyright 2023, American Physical Society. **(e)** Strain effect on optical properties of the $1b_1 \rightarrow 2b_1$ transition at $C_BC_N$ in h-BN as parallel (to $C_2$ symmetry axis, ∥) and perpendicular (⊥) strain is applied. (left) Optical spectra (only the energy range below the bulk-state transition is shown) are red-shifted as the strain increases at both random-phase approximation (RPA) level (blue) and BSE level (orange). The radiative lifetime is denoted as $\tau_R$. (right) BSE excitation energy and ZPL energy (up), exciton binding energy ($E_b$, down) as a function of strain. Reproduced with permission from Zhang *et al.*, 2D Mater. **10**, 035036 (2023)[160]. Copyright 2023, IOP publishing..